 \documentclass[final,authoryear,5p,twocolumn]{elsarticle}

\usepackage{amssymb}
\usepackage{amsmath}
\usepackage{url}

\journal{J. of Atmospheric and Solar-Terrestrial Physics}

\begin{document}

\begin{frontmatter}




\title{ Multi-scale harmonic model for solar and climate cyclical variation throughout the Holocene based on Jupiter-Saturn tidal frequencies plus the 11-year solar dynamo cycle}
%


\author{Nicola Scafetta $^{1}$}

 \address{$^{1}$ACRIM (Active Cavity Radiometer Solar Irradiance Monitor Lab)  \& Duke University, Durham, NC 27708, USA.}
\begin{abstract}
 The Schwabe frequency band of the Zurich sunspot record since 1749   is found to be made of three major  cycles with periods of about 9.98, 10.9 and 11.86 years. The side frequencies appear to be closely related to the spring tidal period of Jupiter and Saturn (range between 9.5-10.5 years, and median 9.93 years) and to the tidal sidereal period of Jupiter (about 11.86 years). The central cycle may be associated to a  quasi 11-year solar dynamo cycle that appears to be approximately synchronized to the average of the two planetary  frequencies. A simplified harmonic constituent model based  on the above two planetary tidal frequencies and on the exact dates of Jupiter and Saturn planetary tidal phases, plus a theoretically deduced 10.87-year central cycle reveals complex quasi-periodic interference/beat patterns. The major beat periods occur at about 115, 61 and 130 years, plus a quasi-millennial large beat cycle around 983 years. We show that equivalent synchronized cycles are found in cosmogenic records used to reconstruct solar activity and in proxy climate records throughout the Holocene(last12,000years) up to now. The quasi-secular beat oscillations hindcast reasonably well the known prolonged periods of low solar activity during the last millennium known as Oort, Wolf, Sp\"orer, Maunder and Dalton minima, as well as   the seventeen 115-year long  oscillations found in a detailed  temperature reconstruction of the Northern Hemisphere covering the last 2000 years.  The millennial three-frequency beat cycle hindcasts equivalent solar and climate cycles for 12,000 years. Finally, the harmonic model herein proposed reconstructs the prolonged solar minima that occurred during 1900-1920 and 1960-1980, the secular solar maxima around 1870-1890, 1940-1950 and 1995-2005, and a secular upward trending during the 20th century: this modulated trending agrees well with some solar proxy model, with the ACRIM TSI satellite composite and with the global surface temperature modulation since 1850. The model forecasts a new prolonged solar grand minimum during 2020-2045, which would be produced by the minima of both the 61 and 115-year reconstructed cycles. Finally, the model predicts that during low solar activity periods, the solar cycle length tends to be longer, as some researchers have claimed. These results  clearly indicate that  solar and climate oscillations are linked to planetary motion and, furthermore, their timing can be reasonably hindcast and forecast for decades, centuries and millennia. The demonstrated geometrical synchronicity between solar and climate data patterns with the proposed solar/planetary harmonic model rebuts a major critique (by Smythe and Eddy, 1977) of the theory of planetary tidal influence on the Sun. Other qualitative discussions are added about the plausibility of a planetary influence on solar activity.
\end{abstract}

%
\begin{keyword}
Planetary theory of solar variation;
Coupling between planetary tidal forcing and solar dynamo cycle;
Reconstruction and forecast of solar and climate dynamics  during the Holocene;
Harmonic model for solar and climate variation at decadal-to-millennial time-scales.\newline\newline
\emph{Cite this article as:} Scafetta, N., Multi-scale harmonic model for solar and climate cyclical variation throughout the Holocene based on Jupiter-Saturn tidal frequencies plus the 11-year solar dynamo cycle. Journal of Atmospheric and Solar-Terrestrial Physics (2012). doi:10.1016/j.jastp.2012.02.016 \newline\newline
\emph{On line version at}: \url{http://dx.doi.org/10.1016/j.jastp.2012.02.016}

\end{keyword}

\end{frontmatter}


%
\section{Introduction}

It is currently believed that solar activity is driven by internal solar dynamics alone. In particular, the observed quasi-periodic 11-year cyclical changes in the solar irradiance and sunspot number,  known as the Schwabe  cycle, are believed to be the result of solar differential rotation as modeled in hydromagnetic solar dynamo models \citep{Tobias,Jiang}.
However, solar dynamo models are not able of hindcasting or forecasting    observed solar dynamics. They fail to properly reconstruct the variation of  the solar cycles that reveal a decadal-to-millennial variation in solar activity \citep{Eddy,Hoyt,Bard,Ogurtsov,Steinhilber}.

 \cite{Ogurtsov} studied the power spectra of multi-millennial solar related records. These authors have found that in addition to the well-known Schwabe 11-year and Hale 22-year solar cycle, there are other important cycles. They found that: (1) an ancient sunspot record based on naked eye observations (SONE) from 0 to 1801 A.D.  presents significant frequency peaks at 68-year and 126-year periods; (2) a $^{10}$Be concentration record in South Pole ice data for 1000-1900 A.D. presents  significant  64-year and 128-year cycles; (3)  a reconstruction of sunspot Wolf numbers from 1100 to 1995 A.D. presents  significant frequency peaks at 60-year and 128-year periods. Longer cycles are present as well.  The 50-140 year band is usually referred to as the Gleissberg frequency band. The Suess/de Vries frequency band of 160-260 years appears to be a superior harmonic of the Gleissberg band.  Periodicities of  11 and 22 years and in the  60-65, 80-90, 110-140,  160-240, $\sim$500, 800-1200 year bands, as well as longer multi-millennial cycles are often reported and also found in Holocene temperature proxy reconstructions \citep{Schulz,Bard,Ogurtsov,Steinhilber,Fairbridge,Vasiliev}.  In particular, \cite{Ogurtsov} also found that mean annual temperature proxy reconstructions in the northern hemisphere from 1000 to 1900 present a quite prominent 114-year cycle, among other cycles.

 None of the above cycles can be explained with the models based on  current mainstream solar theories, probably because  solar dynamics is not determined by internal solar mechanisms alone, as those theories assume, and, further, the physics explaining the dynamical evolution of the Sun is still largely unknown.

An alternative theory has been  proposed and studied since the  19$^{th}$ century  and it was originally  advocated even by well-known scientists such as Wolf (1859), who named the sunspot number series. This thesis was also advocated  by many solar and aurora experts \citep{Lovering} and by other scientists up to now \citep{Schuster,Bendandi,Takahashi,Bigg,Jose,Wood65,Wood,Dingle,Okal,Fairbridge,Charvatova90,Charvatova20,Charvatova09,Landscheidt88,
Landscheidt99,Hung,Wilson,Scafetta,Wolff,Scafettaconf,scafett2011b,Scafetta2011}.

Indeed, the very first proposed solar cycle theory  claimed that solar dynamics could be partially driven by the varying gravitational tidal forces of the planets as they orbit the Sun. However, since the 19$^{th}$ century the theory has also been strongly criticized in many ways. For example, it was found that Jupiter's period of 11.86 years poorly matches the 11-year sunspot cycle: see the detailed historical summary about the rise and fall of the \emph{first solar cycle model} in \cite{Charbonneau}.

In the following we will see how some of the major critiques can be solved.
If planetary tidal forces are influencing the Sun in some way, their frequencies should be present in solar dynamics, and it should be possible to reveal them with high resolution data analysis methodologies. Of course, planetary harmonics would only act as an external forcing that constrains solar dynamics to an ideal  cycle around which the sun chaotically fluctuates. A planetary theory of solar dynamics should not be expected to describe every detail manifested in the solar observations. It can only provide a schematic and idealized representation of such a dynamics.

There are two major planetary tidal frequencies within the Schwabe frequency band. These are the sidereal tidal period of Jupiter (about 12 years) and the spring tidal period of Jupiter and Saturn (about 10 years). If solar activity is partially driven by these two tidal cycles, their frequencies should be found in the solar records. The irregular Schwabe 11-year solar cycle could be the product of an approximate synchronization of the solar dynamics to these two major tidal cycles. However, the solar dynamo cycle too should  actively contribute to produce the final solar dynamic output.

Moreover, if solar dynamics is characterized by a set of quasi-periodic cycles, complex interference and beat patterns should emerge by means of harmonic superposition. Sometimes these cycles may generate constructive interference periods, and sometimes they may give origin to destructive interference periods. Thus, multi-decadal, multi-secular and millennial solar variations may result from the complex interference of numerous superimposed cycles. For example, we may expect that during the periods of destructive interference the Sun may enter into prolonged periods of minimum activity, while during the periods of constructive interference the Sun may experience prolonged periods of maximum activity. If solar dynamics can be partially reconstructed by using a given set of harmonics, forecasting it would also be possible with a reasonable accuracy.

The above results are  also important for climate change research too.  In fact, solar variations have been associated to climate changes at multi time-scales by numerous authors \citep{Eddy,Sonett,Hoyt,Bond,Kerr,Kirkby,Eichler,Soon,Scafetta2007,Scafetta2009,Scafetta}. Thus, reconstructing and forecasting solar changes may be very useful to better understand past climate changes and partially forecast them by projecting  multi-decadal, secular and millennial climate cycles.

  \begin{figure*}[t]
\includegraphics[angle=-90,width=40pc]{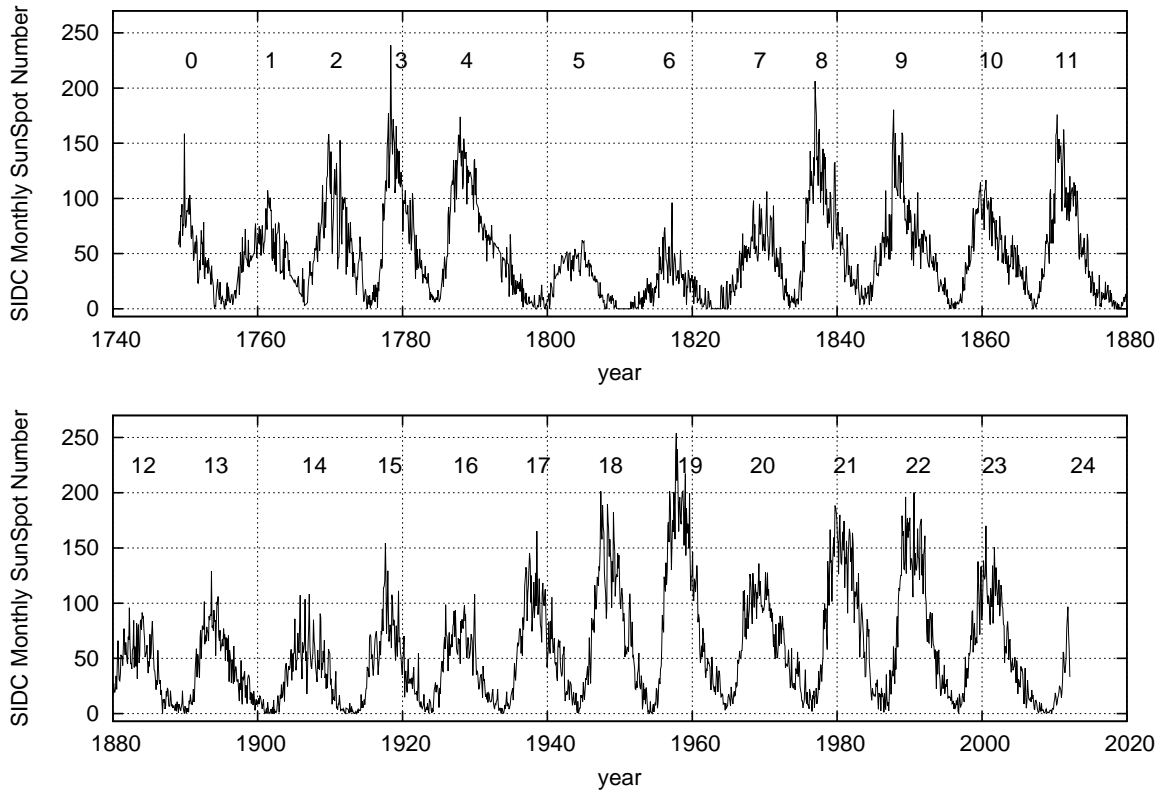}
 \caption{The monthly average sunspot number record from the Solar Influences Data Analysis Center (SIDC) from January/1749 to February/2012. (http://sidc.oma.be/index.php)}
 \end{figure*}

In the following we construct a simplified version of a harmonic constituent solar cycle model based on solar and planetary tidal cycles.
It is worth to note that the harmonic constituent models that are  currently used to accurately reconstruct and forecast tidal high variations on the Earth, are typically  based on observed solar and lunar cycles and made of 35-40 harmonics \citep{Thomson,Ehret}. What we propose here is just a basic simplified prototype of a harmonic constituent model based on only the three major frequencies revealed by the power spectrum of the sunspot number record. We simply check if this simplified model may approximately hindcast the timing of the major observed solar and climate multi-decadal, multi-secular and multi-millennial known patterns, which will tell us   whether planetary influence on the Sun and, indirectly, on the climate of the Earth should be more extensively investigated.

We will explicitly show that our simplified harmonic model suffices to rebut at least one of the major critiques against the plausibility of planetary influence on the Sun that was proposed by Smythe and Eddy (1977). These authors in the early 1980s convinced most solar scientists to abandon the theory of a planetary influence on the Sun. This critique was based on a presumed geometrical incompatibility between the dynamics revealed by planetary tidal forces and the known dynamical evolution of solar activity, which presents extended periods of low activity such as during the Maunder minimum (1650-1715). We will show how the problem can be easily solved.

We also add some qualitative comments concerning recent findings \citep{Wolff,Scafetta2011}, which may  respond to other traditional  critiques such as the claim   that planetary tides are too small to influence the Sun \citep{Jager}.

It is not possible to accurately reconstruct  the \emph{amplitudes} of the solar dynamical patterns. Such exercise would be impossible also because the multi-decadal, multi-secular and millennial amplitudes of the total solar luminosity and solar magnetic activity are extremely uncertain. In fact,  several authors, and sometimes even the same author (Lean), have proposed quite different amplitude solutions \citep{Lean,Hoyt,Bard,Krivova,Steinhilber,Shapiro}. We are only interested in determining whether the frequency and  timing patterns of the known solar reconstructions reasonably match with our proposed model. Indeed, despite that the amplitudes are quite different from model to model, all proposed solar activity reconstructions present relatively similar patterns such as, for example,  the  Oort, Wolf, Sp\"orer, Maunder and Dalton grand solar minima at the corresponding dates.

For convenience of the reader, an appendix and an supplement data file are added to describe the equations and the data of the proposed solar/planetary harmonic model.

  \begin{figure}[t]
\includegraphics[angle=-90,width=21pc]{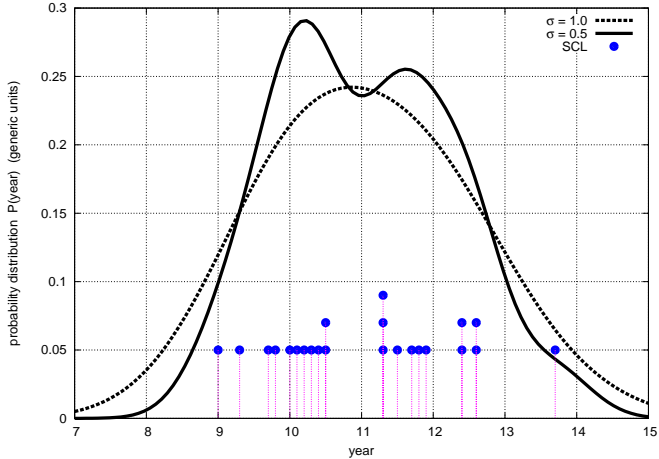}
 \caption{Probability distributions of the sunspot number cycle length (SCL) using Table 1 and Eq. 2. Note the two-belled distribution pattern (solid curve) centered close to 10 and 12 years. There appears to be a gap between 10.55 and 11.25 years in the SCL record. }
 \end{figure}

\section{The 9-13 year Schwabe frequency band is made of three frequencies at about 9.93, 10.9 and 11.86 years}

We use the Wolf sunspot number record from the Solar Influence Data Analysis Center (SIDC): see Figure 1. This record contains 3157 monthly average sunspot numbers from January 1749 to January 2012 and presents 23 full Schwabe solar cycles, whose length is calculated from minimum to minimum.  Table 1 reports the approximate starting and ending dates of the 23 solar cycles and their approximate amplitude and length in years.

Figure 2 shows two probability distributions, $P(x)$, of the lengths of the 23 observed  Schwabe  cycles. Because of the paucity of the data, the probability distributions are estimated by associating to each cycle $i$ with length $L_i$ a normalized Gaussian probability function of the type

\begin{equation}\label{Eq2}
    p_i(x)=\frac{1}{\sqrt{2\pi\sigma^2}} ~ \exp\left[-\frac{(x-L_i)^2}{2\sigma^2}\right]~,
\end{equation}
where $\sigma$ is a given standard deviation.
Then, the probability distribution of all cycles is given by the average of all singular distributions as:

\begin{equation}\label{Eq3}
    P(x)=\frac{1}{23}\sum_{i=1}^{23} p_i(x)~.
\end{equation}
The choice of $\sigma$ depends on the desired smoothness of the probability distribution.
If $\sigma=1$ is used, a single belled shape distribution appears. This distribution spans from 9 to 13 years, which can be referred to as the Schwabe frequency band. Its maximum is at about 10.9 years. Note that the bell is not perfectly symmetric but skewed toward longer cycles: the average among the 23 cycle lengths is 11.06 years.

\begin{table}[t]
\begin{tabular}{c c c c c c}
  \hline
SC	&	Started	&	SCMin	&	SCL	& SCMY & SCMax \\ \hline
1	&	03/1755	&	8.4	&	11.3    &   06/1761    &  86.5 	  \\ \hline
2	&	06/1766	&	11.2&	9       &	09/1769    & 115.8 \\ \hline
3	&	06/1775	&	7.2	&	9.3     &	05/1778    & 158.5  \\ \hline
4	&	09/1784	&	9.5	&	13.7    &	10/1787    & 138.1  \\ \hline
5	&	05/1798	&	3.2	&	12.6    &	02/1804    &  49.2  \\ \hline
6	&	12/1810	&	0.0	&	12.4    &	05/1816    &  48.7  \\ \hline
7	&	05/1823	&	0.1	&	10.5    &	07/1830    &  70.7  \\ \hline
8	&	11/1833	&	7.3	&	9.8     &	03/1837    & 146.9 \\ \hline
9	&	07/1843	&	10.6&	12.4    &	02/1848    & 131.9 \\ \hline
10	&	12/1855	&	3.2	&	11.3    &	02/1860    &  98.0   \\ \hline
11	&	03/1867	&	5.2	&	11.8    &	08/1870    & 140.3 \\ \hline
12	&	12/1878	&	2.2	&	11.3    &	12/1883    &  74.6 \\ \hline
13	&	03/1890	&	5.0	&	11.9    &	01/1894    &  87.9 \\ \hline
14	&	02/1902	&	2.7	&	11.5    &	02/1906    &  64.2 \\ \hline
15	&	08/1913	&	1.5	&	10      &	08/1917    & 105.4 \\ \hline
16	&	08/1923	&	5.6	&	10.1    &	04/1928    &  78.1 \\ \hline
17	&	09/1933	&	3.5	&	10.4    &	04/1937    & 119.2  \\ \hline
18	&	02/1944	&	7.7	&	10.2    &	05/1947    & 151.8 \\ \hline
19	&	04/1954	&	3.4	&	10.5    &	03/1958    & 201.3 \\ \hline
20	&	10/1964	&	9.6	&	11.7    &	11/1968    & 110.6 \\ \hline
21	&	06/1976	&	12.2&	10.3    &	12/1979    & 164.5 \\ \hline
22	&	09/1986	&	12.3&	9.7     &	07/1989    & 158.5 \\ \hline
23	&	05/1996	&	8.0	&	12.6    &	04/2000    & 120.8 \\ \hline
24	&	12/2008	&	1.7	&    &	&	\\ \hline
\end{tabular}
\caption{Starting and ending approximate dates of the 23 observed Schwabe sunspot cycle since 1749. The table also reports the approximate solar cycle length (SCL) in years, its annual average minimum (SCMin) and maximum (SCMax) and the year of the sunspot number maximum (SCMY). The average length of the 23 solar cycles is 11.06 years. }\label{tb1}
 \end{table}

If $\sigma=0.5$ is used, two distribution peaks appear close to about 10 and 12 years. This double belled distribution is physically interesting because it reveals that the solar cycle dynamics may be constrained by two major frequency attractors at about 10 and 12 year periods, respectively. Thus, the solar cycle length does not appear to be just a random variable distributed on a single-belled Gaussian function centered around an 11-year periodicity (as typical solar dynamo models would predict), but it appears to be generated by a more complex dynamics driven by two cyclical side attractors.

  \begin{figure*}[t]
\includegraphics[angle=0,height=30pc,width=40pc]{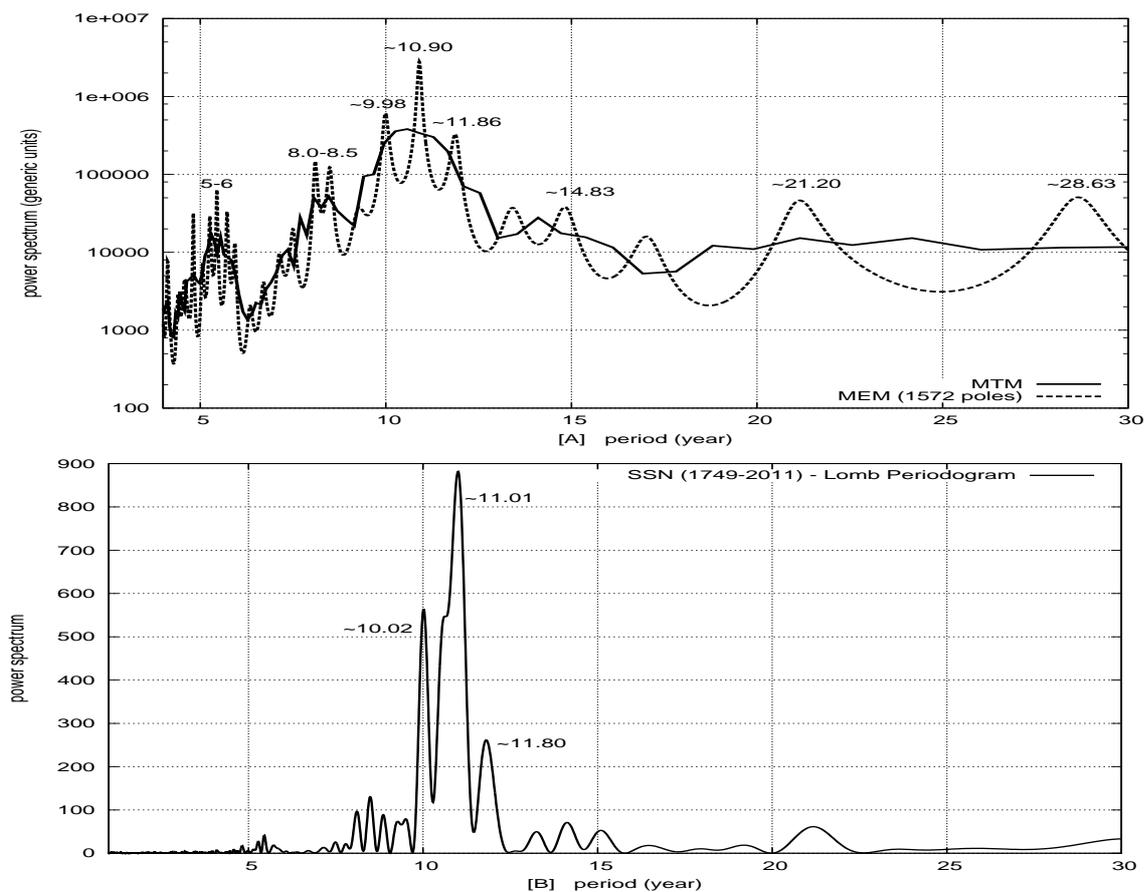}
 \caption{Power spectrum analysis of the monthly average sunspot number record. [A] We use the multi taper method (MTM) (solid) and the maximum entropy method (MEM) (dash) (Ghil et al., 2002). Note that the three frequency peaks that make the Schwabe sunspot cycle at about 9.98, 10.90 and 11.86 years are the three highest peaks of the spectrum. [B] The Lomb-periodogram too reveals the existence of  the  three spectral peaks.}
 \end{figure*}

The bottom of Figure 2 shows in circles the 23 actual sunspot cycle lengths used to evaluate the two distributions. In our evaluation, no Schwabe solar cycle with a length between 10.55 and 11.25 years is observed. The existence of this gap reinforces the interpretation that solar cycle dynamics may be driven by two dynamical attractors with periods at about 10 and 12 years. In fact, the area below the single-belled distribution curve within the interval 10.55 and 11.25 is 0.17. It is easy to calculate that the probability to get by  chance 23 random consecutive measurements from the single-belled probability distribution depicted in Figure 2 outside its central 10.55-11.25 year interval is just $P=(1-0.17)^{23}=1.4\%$, which is a very small probability. Thus, the solar cycle length does not appear to be just a random variable of a single-belled distribution.   Paradoxically, we may conclude that a rigorous 11-year Schwabe solar cycle does not seem to exist in solar dynamics or, at least, it has never been observed since 1749. The 11-year period is just an ideal average cycle. The bimodal nature of the solar cycle has been noted also by other scientists \citep{Rabin,Wilson11}.

Figure 3 shows high resolution spectral analysis of the monthly resolved sunspot number record from Jan/1949 to Dec/2010 for 3144 monthly data points. The multi taper method (MTM) (solid line) indicates that the sunspot number record presents a wide Schwabe peak with a period spanning approximately from 9 to 13 years, as seen in the solar cycle length probability distribution depicted in Figure 2. The second analysis uses the maximum entropy method (MEM) \citep{Press}. The third analysis uses the traditional Lomb-periodogram, which further confirms the results obtained with MEM. We use MEM with the highest allowed pole order, which is half of the length of the record (1572 poles): see \cite{Courtillot1977} and \cite{scafett2012} for extended discussions on how to use MEM.

     The spectral analysis reveals that the wide sunspot 9-13 year frequency band is made of three frequencies with periods at about 9.98 year (which is compatible with the 9.5-10.5 year Jupiter/Saturn tidal spring period), the central 10.90 year (the top-distribution Schwabe solar cycle, see Figure 2 dash curve), and 11.86 year (which is perfectly compatible with the 11.86-year sidereal period of Jupiter). By adding or subtracting 1 year from the data record, the periods associated to these three spectral peaks vary less than 0.1 year.  Thus, the spectral analysis appears to   resolve the two attractor frequencies close to 10 and 12 years revealed by the solar cycle length distribution depicted in Figure 2. These three peaks are the strongest of the spectrum, two of which can be clearly associated to physical planetary tidal cycles, and also generate clear harmonics in the 5-6 year period band.

Note that Figure 3 shows three major cycles while Figure 2 argues about two side-attractors. This is not a contradiction. The ``solar cycle length'' used in Figure 2 is a kind of Poincar\'e section (that is a dynamical sub-section) of the sunspot number dynamics used to produce Figure 3. Thus, the two observables are different; although the former is included in the latter, but not vice versa. Essentially, the two side attractors at about 10 and 12 year periods would force the solar cycle length to be statistically shorter or longer than 11 years by making the occurrence of a rigorously 11-year solar cycle length unlikely. On the contrary, when the sunspot number record is directly analyzed, the central 11-year cycle dominates because it characterizes the \emph{amplitude} of the Schwabe cycle, which appears to be mostly driven by  solar dynamo cycle mechanisms.

Other researchers have studied the fine structure in the sunspot number spectrum with MEM. Currie (1973) found a double solar cycle line in the Zurich sunspot time history from 1749 to 1957 at about 10 and 11.1 years. Currie's figure 3 shows that the peak at 11.1 years is quite wide and skewed toward the 12-year periodicity. Thus, it was likely hiding an unresolved spectral split.
Kane (1999) found for the period 1748-1996 three peaks at 10, 11 and 11.6 years.
Thus, our result that the Schwabe cycle is made of three lines at about 9.98, 10.90 and 11.86 years, is consistent  and can be considered an improvement of  Currie and Kane's results since  statistical accuracy increases with the number of available data.

If circular orbits and constant orbital speeds are assumed, the tidal spring period of Jupiter and Saturn is

\begin{equation}\label{}
    P_{JS}=\frac{1}{2}~\frac{P_J P_S}{P_S-P_J} = 9.93~yr,
\end{equation}
where $P_J=11.862$ year and $P_S=29.457$ year are the sidereal periods of Jupiter and Saturn, respectively: the tidal spring period is half of the synodic period because the tides will be the same whether the planets are aligned along the same side or in opposition relative to the Sun. However, an estimate of the J/S spring period based on the JPL Horizons ephemerides orbital calculations from 1750 to 2011, gives a value oscillating between 9.5 and 10.5 years. This range and its average, $P_{JS}=9.93~yr$, agrees well with the measured $9.98$ year spectral cycle within only 18 days, which is less than  the monthly resolution of the sunspot number record.
Figure 3 also shows other frequency peaks at about 5-6 years, 8-8.5 years, 14.5-15 year, 21-22 years (that is the Hale cycle) and 28-29 years, but these cycles are less important than the three major Schwabe peaks observed above, and their contribution and origin are left to another study.

Indeed, Jupiter and Saturn are not the only planets that may be acting on the Sun.
Also the terrestrial planets (Mercury, Venus and  Earth) can produce approximately equivalent  tides on the Sun \citep{Scafetta2011}. In particular, the Mercury/Venus orbital combination repeats almost every 11.08 years. In fact, $46P_M=11.086~yr$ and $18P_V=11.07~yr$, where $P_M=0.241~yr$ and $P_V=0.615~yr$ are the sidereal periods of Mercury and Venus, respectively. The orbits of Mercury and Venus repeat every 5.54 years, but the 11.08 year periodicity would better synchronize with the orbits of the Earth and with the tidal cycles of the Jupiter/Saturn sub-system.  In any case, there are tidal resonances produced by Mercury and Venus at 5.54 years and 11.08 years. This property may also explain the 5-6 year large frequency peak observed in Figure 3A, which, however, may also be a harmonic of the 10-12 year cycles. In the same way, it is easy to calculate that the orbits of the Earth ($8P_E=8~yr$) and of Venus ($13P_V=7.995~yr$) repeat every 8 years, which may explain the other spectral peak at about 8-year period. Finally,  the tidal patterns produced by  Venus, Earth and Jupiter on the Sun form cycles with a periodicity of 11.07 years, which are also  synchronized to the Schwabe's cycles  \citep{Bendandi,Wood,Hung,Scafetta2011}. Indeed, the 11.07-year resonance period is very close to the measured 11.06-year average sunspot cycle length (see Table 1). Thus, all major frequency peaks observed in Figure 3 may be associated to planetary tides and may explain why the solar dynamics is synchronized around an 10-12 year cycle.

  \begin{figure*}[t]
\includegraphics[angle=0,height=30pc,width=40pc]{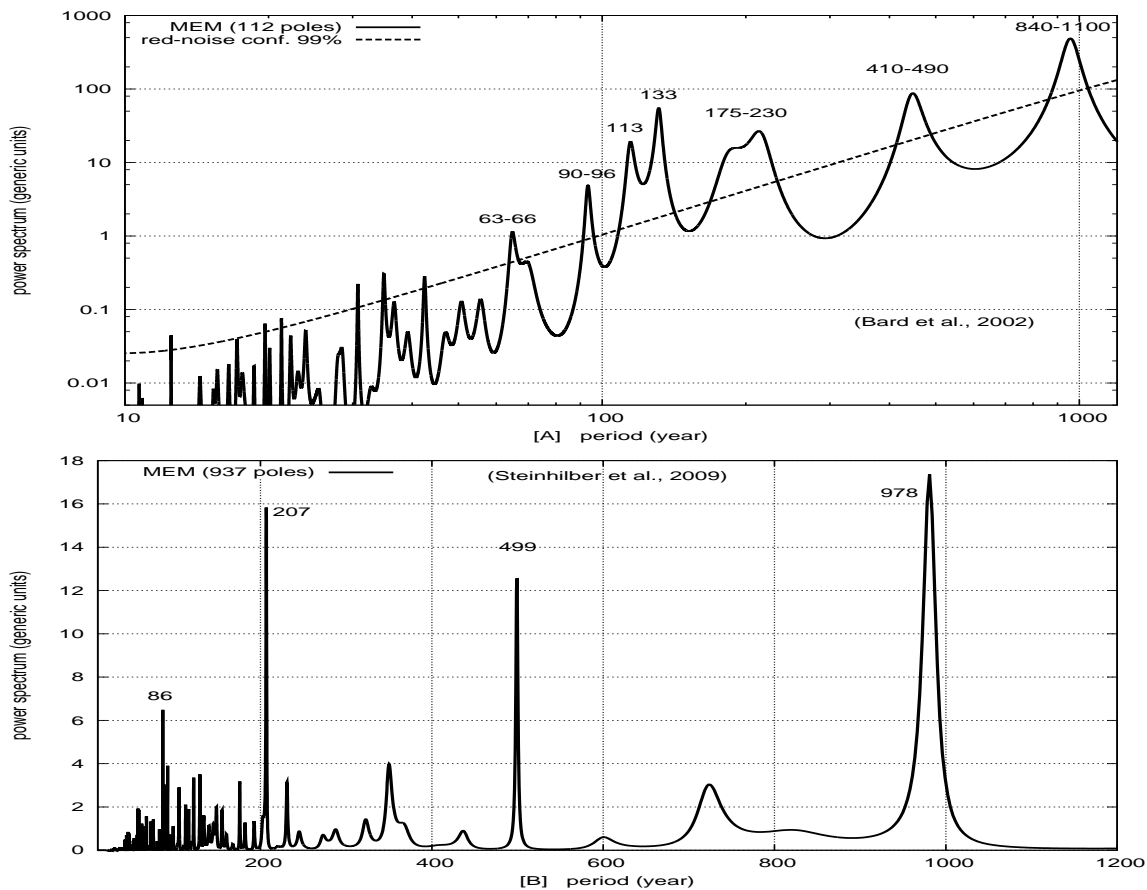}
 \caption{Power spectrum estimates of two total solar irradiance proxy reconstructions by \cite{Bard} (period from 843 A.D. to 1961 A.D.) and by \cite{Steinhilber} (period from 7362 B.C. to 2007 A.D.), respectively. }
 \end{figure*}

In Section 4 we  also use two total solar irradiance proxy reconstructions that were proposed by Bard et al. (2000) (period from 843 A.D. to 1961 A.D.) and  by \cite{Steinhilber} (period from 7362 B.C. to 2007 A.D.) together with other solar and climate records \citep{Bond,Ljungqvist}.  Figure 4 shows the power spectrum estimates associated to these two records. Figure 4A, which refers to \cite{Bard},  shows major peaks at periods of  63-66, 90-96, 113-133, 175-230, 410-490 and 840-1100 years. Figure 4B, which refers to \cite{Steinhilber}, confirms the low frequency peaks, in particular a quasi-millennial cycle of about 978 years. Note that for the latter solar reconstruction, the higher frequencies (corresponding to periods shorter than 200 years) are likely a noise because of the model uncertainty that greatly increases toward the past.      The 50-140 year Gleissberg frequency band, the 160-260 year Suess frequency band, and the quasi millennial solar cycle were found by several authors (see the Introduction), and are confirmed by our results depicted in Figure 4. In conclusion, we have that the 9-13 year Schwabe frequency band is made of three frequencies at about 9.93, 10.9 and 11.86 years within an error of 0.1 years; we will use these three frequencies to build our model.

\section{Analysis of the beats of the three frequencies measured in the Schwabe's frequency band}

Figure 3A indicates the existence of the three major cycles with periods: $P_1=9.98$, $P_2=10.90$ and $P_3=11.86$ years. We observe that the superpositions of these three cycles generate four major beat cycles.  If $P_i$ and $P_j$ are the periods of the two cycles, the beat period $P_{ij}$  is given by

\begin{equation}\label{Eq45}
    P_{ij}=\frac{1}{1/P_i-1/P_j}.
\end{equation}
Assuming a 0.5\% error, which would correspond to about one month resolution of the sunspot record herein used, we have the following beat cycles: $P_{13}=63\pm3$ years; $P_{12}=118\pm10$ years;  $P_{23}=135\pm12$ years. Finally, there is a three-frequency quasi-millennial beat cycle at
\begin{equation}\label{Eq46}
    P_{123}=\frac{1}{1/P_1-2/P_2+1/P_3}\approx 970~yr.
\end{equation}
 In the latter case, the error distribution is quite skewed, and   the result is quite sensitive to the error of the original frequencies. Note that a quasi-millennial cycle could also be forced on the Sun by the rotation of the \emph{Trigon} of the great conjunctions of Jupiter and Saturn \citep{Masar,Kepler2,scafett2011b}.

Thus, the three-frequency model is able to produce beat cycles at about 60-66 years, 108-128 years, 123-147 years, and quasi-millennial cycles with approximate period of 840-1100 years. These frequency ranges (and their harmonics) agree well with many of the major solar frequencies found by numerous authors and in Figure 4. This fact suggests that the oscillations found in the Gleissberg, Suess and the millennial frequency bands could be induced by beats in the three major Schwabe frequencies herein studied. In the next section we optimize these cycles.

\section{The three-frequency solar-planetary harmonic model based on the two Jupiter-Saturn tidal harmonics plus the central Schwabe cycle}

Our proposed three-frequency harmonic model is simply made of  the function:

\begin{equation}\label{Eq5}
    f_{123}(t)=\sum_{i=1}^3 A_i ~ \cos\left(2\pi~\frac{t-T_i}{P_i}\right).
\end{equation}
See the Appendix for the full set of equations. For simplicity, we  set the above function to zero when $f(t)<0$ to approximately simulate the sunspot number record, which is positive defined: see Eq. \ref{Eq14} in the Appendix.

The parameters of our model are chosen under the assumption that the measured side frequencies at approximately 9.98 and 11.86 years correspond to the two major tidal frequencies generated by Jupiter and Saturn.

Thus, we use the exact sidereal orbit periods of the two planets ($P_J=11.862242~yr$ and $P_S=29.457784~yr$) to calculate the following three frequencies: $P_1=9.929656~yr$, $P_2=10.87~yr$ and $P_3=11.862242~yr$: we are ignoring possible orbital perturbations, and using orbital data from http://nssdc.gsfc.nasa.gov/planetary/factsheet/

 The estimated value of $P_2$, which is compatible with the central Schwabe frequency measure of Fig. 3, has been calculated using Eq. \ref{Eq46} under the assumption that $P_{123}\approx970$, as deduced above. Note that a value of $P_2=10.87 \pm 0.01~yr$ would produce a millennial 3-frequency beat cycle with a period   between 843 and 1180 years, which is perfectly compatible with the 900-1100 year range period found in Holocene solar proxy records  \citep{Bond}. Moreover, the central frequency at $P_2=10.87~yr$  is slightly larger than  the average frequency period, $2~P_J~ P_{SJ}/(P_J+P_{JS})=10.810255~yr$, which perhaps may be due to a slight perturbation  induced by  other planetary tides, their exact spatial orientation and/or by the differential rotation of the Sun (note that the period difference is about $0.06~yr = 22 ~day$, which is within the solar rotation period of on average about 27 days). Solving this issue is left to another dedicated study. In any case, it may be reasonable to assume that the period of 10.87 year could emerge by means of a resonance/synchronization of the solar dynamo cycle to the two planetary tides. Indeed,   solar dynamo models can predict a central Schwabe period of about 10.8 years \citep{Jiang}: therefore, with small adjustments of the model parameters, solar dynamo theory can likely predict a central 10.87-year cycle too.

With the chosen three frequencies, we get the following four solar/planetary beat frequencies: $P_{12}= 114.783~yr$, $P_{13}= 60.9484~yr$, $P_{23}= 129.951~yr$ and $P_{123}= 983.401~yr$. The relative amplitudes of the harmonics of the model are chosen proportional to the peaks deduced from the sunspot power spectrum of figure 3B. This gives the following normalized values: $A_1=24/29=0.83$, $A_2=29/29=1$, $A_3=16/29=0.55$.

The third important component of the model is the phase timing of the three harmonics.
We use:  $T_1=2000.475$ (the synodic conjunction date of Jupiter and Saturn, 23/June/2000 relative to the Sun,  when the spring tide should be stronger); $T_2=2002.364$ (which is calculated through regression of the harmonic model against the sunspot number record by keeping all other parameters fixed); and $T_3=1999.381$ (the perihelion date of Jupiter,  20/May/1999,  when Jupiter's induced tide is stronger).

  \begin{figure*}[t]
\includegraphics[angle=0,height=30pc,width=40pc]{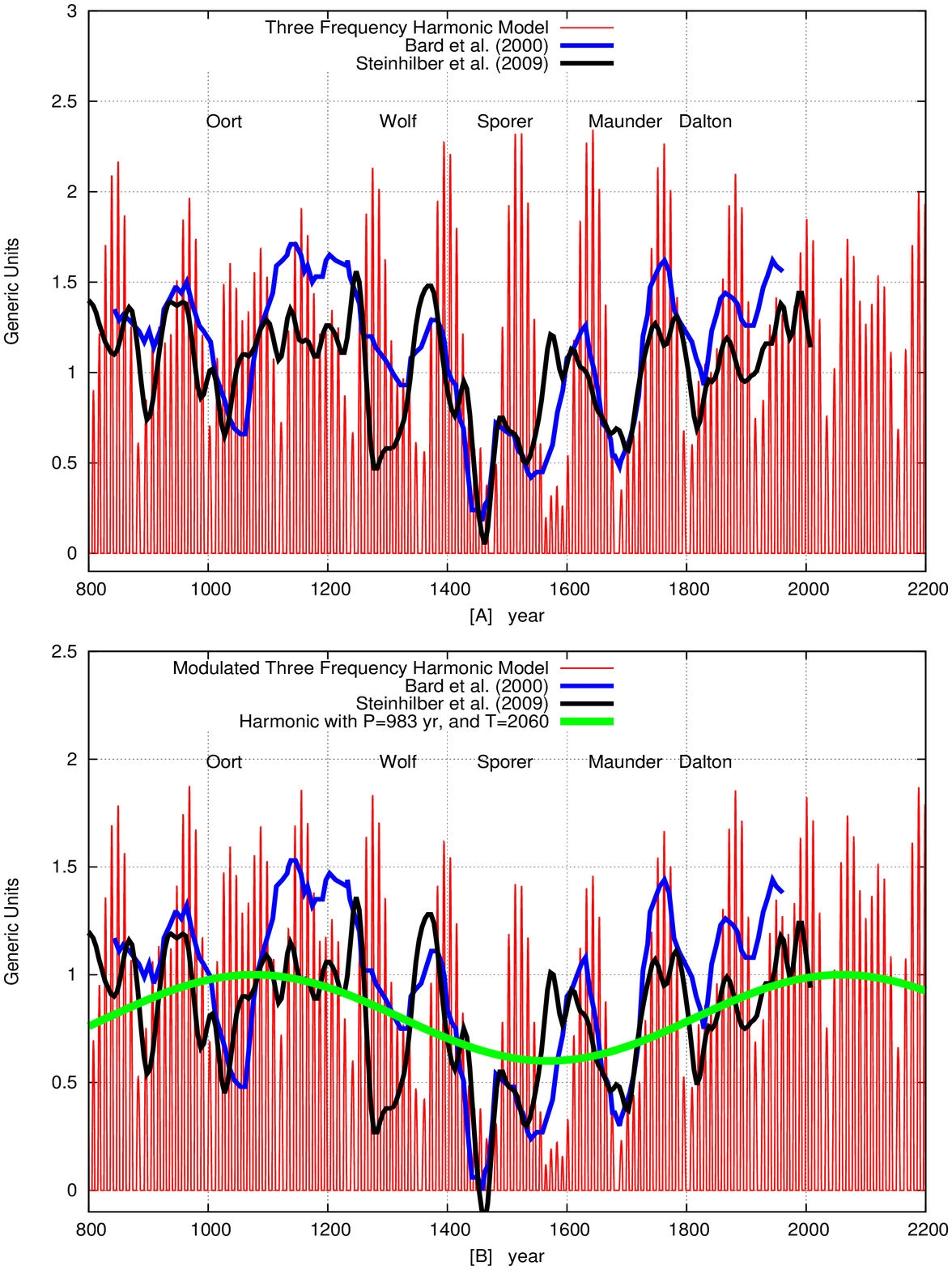}
 \caption{A simple harmonic constituent model based on the three frequencies of the Schwabe sunspot cycle: [A] Eq. \ref{Eq5}; [B] Eq. \ref{Eq555}.   The figures also depict two reconstructions of solar activity based on $^{10}$Be and $^{14}$C cosmogenic isotopes \citep{Bard,Steinhilber}. The millennial harmonic Eq. \ref{Eq5552} is also depicted in [B].  The solar/planetary model show clear  quasi-periodic multi-decadal beat cycles of low and high  activity  with a period of about 110-120 years and a quasi millennial beat solar cycle, which  are  well synchronized to the equivalent cycles observed in the solar reconstructions. The units are generic, but would correspond to $W/m^2$ at 1 AU for Steinhilber's solar reconstruction. }
 \end{figure*}

The phases of the beat functions can be calculated using the following equation:
\begin{equation}\label{9999}
    T_{12}=\frac{P_2T_1-P_1T_2}{P_2-P_1}~.
\end{equation}
We find: $T_{12}=2095.311$, $T_{13}=2067.044$, $T_{23}=2035.043$ and $T_{123}=2059.686$, where $T_{123}$ is calculated by beating $P_{12}$ and $P_{23}$ and subtracting half of the period $P_{123}$ for reasons explained in the next section. Note that the phases for the conjunction of Jupiter and Saturn vary by a few months from an ideal average because of the elipticity of the two orbits, which would imply a variation up to a few years about the phases of the beat functions: herein we ignore these corrections.

In conclusion, on nine parameters that make the proposed combined solar/planetary model, the three amplitudes come from the sunspot number records, two frequencies and their two phases come from Jupiter and Saturn orbital information, and the third frequency and its phase come from a combination of the information deduced from planetary orbits and sunspot numbers since 1749. Perhaps, also the $P_2$ frequency may be deduced from planetary physics without explicitly involving internal solar mechanisms by taking into account all planetary tides and the elipticity of the orbits, but we leave the solution of this more advanced problem to another paper.

\section{Multi-scale hindcast and forecast of solar and climate  records throughout the Holocene}

The proposed model needs to be tested by checking whether the time series built from it, once  extended far before 1749 A.D. (in fact, we are using the sunspot record since 1749 for estimating some of the parameters of the model), is able to approximately hindcast the timing of the major known solar secular patterns and the quasi millennial cycle. Some of the secular low solar activity patterns are named as  Oort, Wolf, Sp\"orer, Maunder and Dalton grand solar minima.  We also need to test whether the solar/planetary model is able to hindcast major secular and millennial patterns observed in the multi-millennial multi-proxy temperature records, which may also be good proxies for solar activity at multiple temporal scales \citep{Bond,Kirkby,Scafetta2007,Scafetta2009}.

We use Eq. \ref{Eq5} to prepare a monthly sequence from 10000 B.C. to 3000 A.D.: see also Eq. \ref{Eq14}. Figure 5A shows the model against two reconstructions of the total solar irradiance since 800 A.D. based on $^{10}$Be and $^{14}$C cosmogenic isotopes \citep{Bard,Steinhilber}.

Figure 5A shows that the solar reconstructions present clear quasi-periodic, secular cycles of low and high solar activity at quasi regular periods of about 110-120 years plus a quasi-millennial solar cycle. These multi-decadal periods of low and high solar activity are well synchronized to the beat patterns produced by the three-frequency harmonic model.
Thus, our simple solar/planetary harmonic model based on just the three major Schwabe solar cycle frequencies is able to predict, with the accuracy of one or two Schwabe solar cycles uncertainty, the timing of all major periods of solar multi-decadal grand minima observed during the last 1000 years. These minima occurred approximately when the three harmonics interfered destructively. On the contrary,  multi-decadal grand maxima of solar activity occurred when the three harmonics interfered constructively.

The model shows also a millennial beat modulation, like the data. We may also assume that this millennial cycle modulates the amplitude of the model signature.  The idea is that when the solar frequencies strongly interfere with each other producing fast changes in the basic harmonic model, the Sun may not to linearly follow those oscillations because of its thermal inertia, and the resulting chaotic state would continuously disrupt solar dynamics preventing the Sun from properly resonating with the harmonics. Thus, we may assume that during these periods, as it happened during the Little Ice Age from 1300 to 1800, the overall millennial-scale solar activity modulation is weakened.   On the contrary, when the interference patterns produce slower changes, resonance dynamical states would be favored yielding a higher average solar activity. Thus, we propose that a more physical harmonic model may contain a non linear coupling between the three-frequency linear model and the millennial cycle produced by the combined beats such as in

  \begin{figure*}
\includegraphics[angle=0,height=43pc,width=40pc]{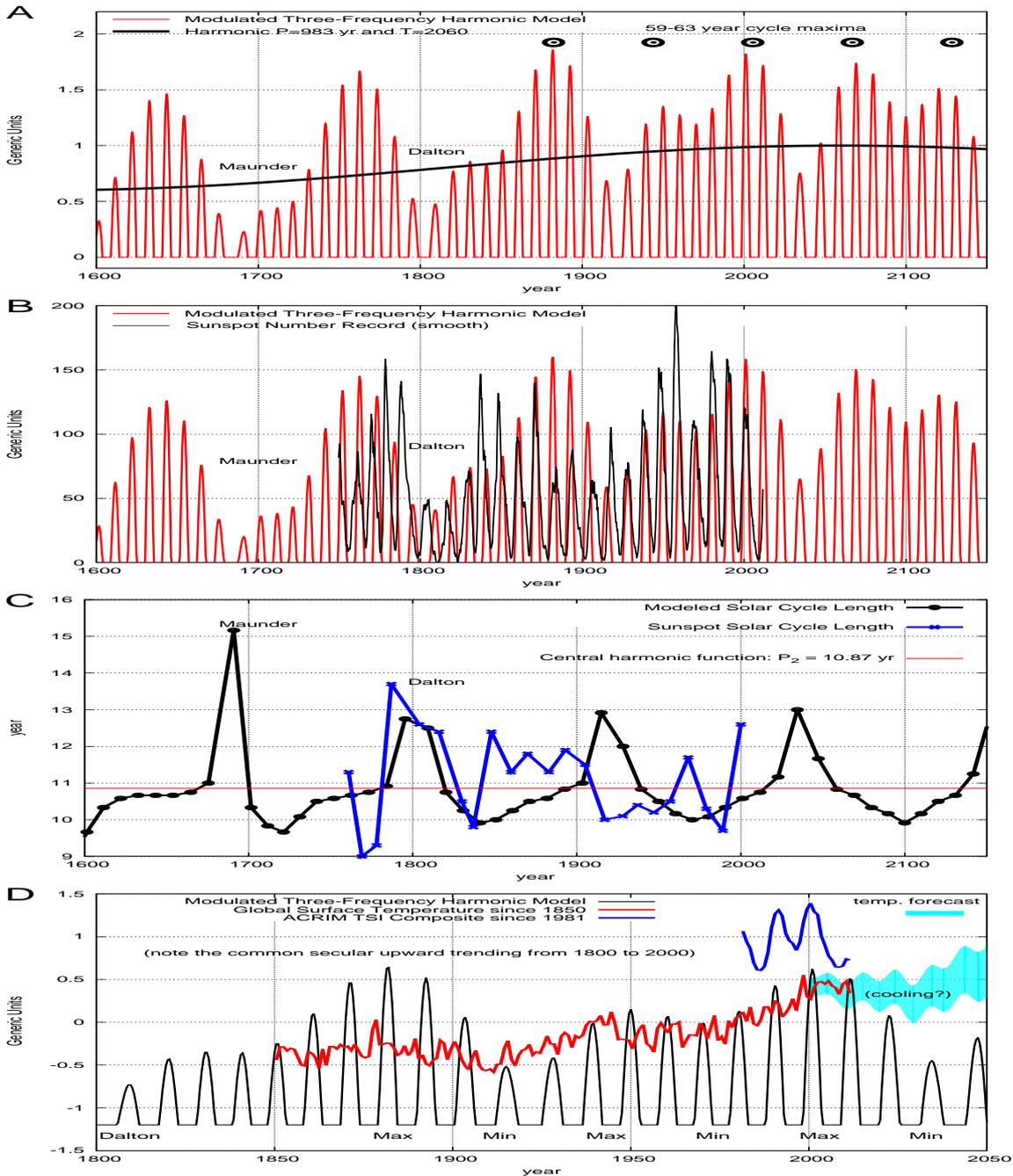}
 \caption{[A] Zoom of Figure 5B from 1600 to 2150. The Maunder Minimum (MM) (1645-1715) and the Dalton Minimum (DM) (1790-1830); the 983-year long solar cycle that peaks in 2060; an upward solar trending from about 1900 to 2000 due mostly to the 115-year cycle; the emergence of five 59-63 year cycles from 1850 to 2150. Note that the next sexagesimal solar minimum should occur around  2025-2043 and the next sexagesimal solar maximum around 2062. [B] Solar planetary model cycle prediction against the sunspot number record: the agreement is approximate as explained in the text. [C] Solar cycle length predicted by the harmonic model against the observed one (see Table 1). Note that the model usually predicts longer solar cycles during periods of lower solar activity as it has been observed during the Maunder and Dalton minima. [D] The solar/planetary model against the global surface temperature (HadCRUT3, http://www.cru.uea.ac.uk/) and the annual average ACRIM total solar irradiance (TSI) satellite composite (http://acrim.com/) since 1981 (note that ACRIM1 started in February/1980). Note the schematic similitude of the modulation of the patterns with local maxima around 1880, 1940 and 2000, local minima around 1910 and 1970, and the similar upward trending since the Dalton solar grand minimum.The preliminary temperature forecast (yellow) area is made with the model proposed in Scafetta (in press) plus a 115-year cycle with amplitude 0.1 $^oC$.}
 \end{figure*}

\begin{equation}\label{Eq555}
    F_{123}(t)\propto g_m(t)~ f_{123}(t).
\end{equation}
More complicated models may be proposed, but here we would like to use only the simplest reasonable case. Figure 5B depicts the above modulated model adopting the following  modulating function:

\begin{equation}\label{Eq5552}
    g_m(t)=0.2~\cos\left(2\pi~\frac{t-2059.7}{983.4}\right)+0.8,
\end{equation}
which has been empirically chosen to approximately reproduce the quasi-millennial beat modulation. The value of the phase, $T_{123}\approx2059.7$  A.D., has been estimated using Eq. \ref{9999}. The chosen amplitude, 0.2, is about the amplitude of the millennial cycle measured in the solar proxy by \cite{Steinhilber} as shown in the figure. However, as explained above, the exact amplitude is uncertain because of the uncertainty among the proposed solar proxy models, and the used amplitude value is hypothetical.

Note that nucleotide proxy models may present large timing error \citep{Bond,Bender}; in fact, the two adopted solar proxy models do present some evident difference from each other.
We may say that the solar/planetary harmonic model has been sufficiently successful in reconstructing major solar patterns during the past 1000 years within the precision of the proxy solar models.

Figure 6 zooms Figure 5B during the period 1600-2150.
 Figure 6A  depicts the Maunder  and Dalton grand solar minima quite well;  a known solar minimum that occurred during the period 1900-1920; reconstructs a known increasing solar activity since 1900-1920 until about 2001 \citep{Scafetta2009,ScafettaW2009}; reconstruct a likely solar maximum during the period 1940-1950, and predicts a new prolonged solar minimum during the period 2020-2050. The emergence of four 59-63 year cycles from 1850 to 2150 is highlighted in the figure with black circles with maxima around 1879, 1940, 2001, 2062 and 2123: note that the first 3 dates correspond well to well-known quasi 60-year temperature maxima \citep{Scafetta,scafett2011b,scafett2012}. The next sexagesimal cycle solar minimum will occur around 2033. The model would qualitatively suggest that there will be an approaching  2025-2043 solar grand minimum and that it will be quite deep, approximately as it was in 1900-1920, because the approaching minimum combines the minima of the 60-year and the 115-year cycles.

  \begin{figure*}[t]
\includegraphics[angle=-90,width=40pc]{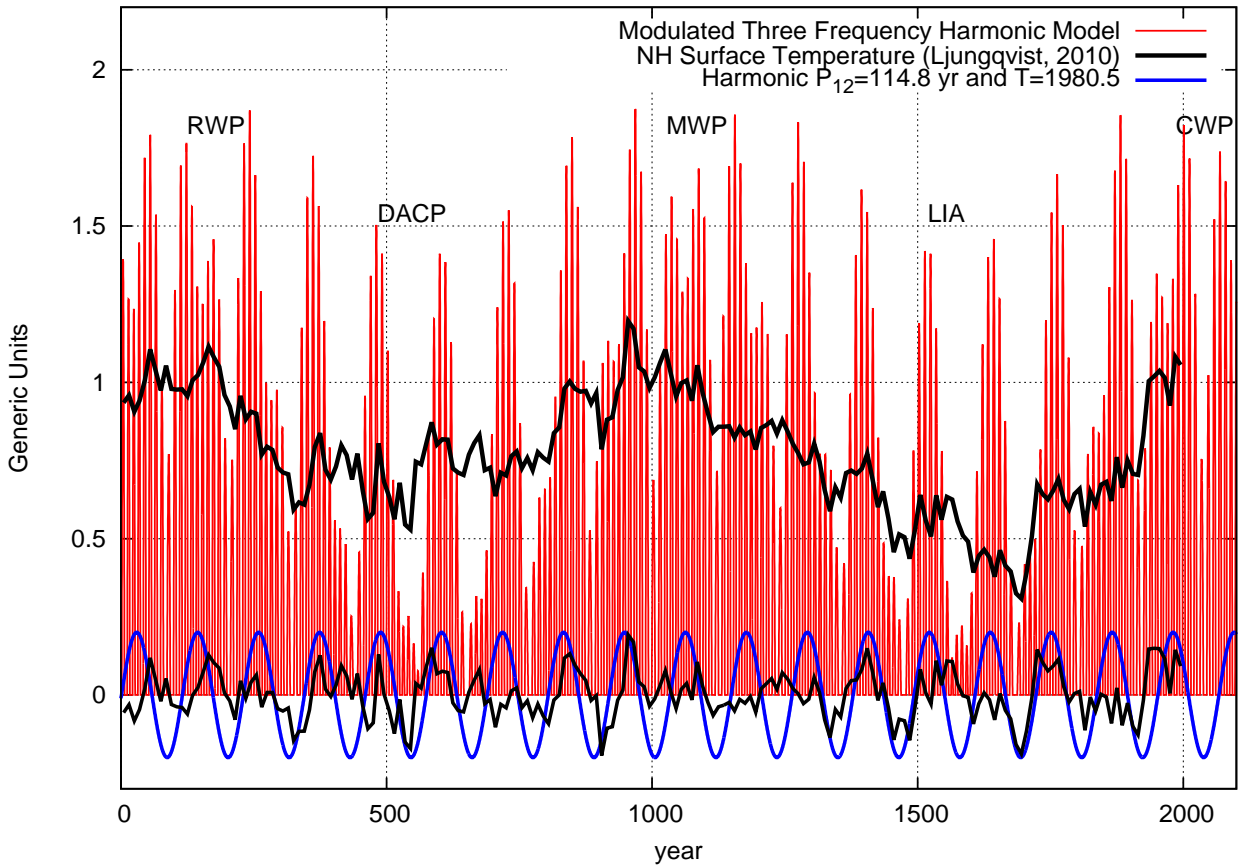}
 \caption{Modulated three-frequency harmonic model, Eq. \ref{Eq555}, (which represents an ideal solar activity variation) versus the Northern Hemisphere proxy temperature reconstruction by \cite{Ljungqvist}. Note the good timing matching of the millenarian cycle and the seventeen  115-year cycles between the two records. The Roman Warm Period (RWP), Dark Age Cold Period (DACP), Medieval Warm Period (MWP), Little Ice Age (LIA) and Current Warm Period (CWP) are indicated in the figure. At the bottom: the model harmonic (blue) with period $P_{12}=114.783$ and phase $T_{12}=1980.528$ calculated using Eq. \ref{9999};  the 165-year smooth residual of the temperature signal. The correlation coefficient is $r_0=0.3$ for 200 points, which indicates that the 115-year cycles in the two curves are very well correlated ($P(|r|\geq r_0)<0.1\%$). The 115-year cycle reached a maximum in 1980.5 and will reach a new minimum in 2037.9 A.D.}
 \end{figure*}

 Figure 6B  depicts the model against the sunspot number record. The Schwabe cycles are approximately recovered. However, the matching is not  perfect, as expected. In fact, the harmonic model uses only three frequencies, while Figure 3 clearly indicates that other frequencies are necessary for an accurate reconstruction of the sunspot number record. There might also be the possibility that the harmonic model represents a real aspect of solar dynamics which is simply not fully reproduced by the sunspot record. For example, the luminosity cycle may not always exactly mirror the sunspot cycle in its amplitude or length or timing. These issues are still open because accurate satellite measurements of solar irradiance began in 1978 \citep{ScafettaW2009}, and we may dedicate another paper to address them.

Figure 6C depicts the modeled solar cycle length that has been calculated from minima to minima of Eq. \ref{Eq13} (see the Appendix). As explained above, because the sunspot record is characterized by more than three frequencies, we should have expected that the agreement is not perfect, as the figure shows. However, the harmonic model usually predicts longer Schwabe cycles between 12 and 15 years, during prolonged solar minima such as during the Maunder and Dalton minima. Indeed, an approximate inverse relation between solar cycle length and average solar luminosity or activity has been postulated  numerous times \citep{Christensen,Loehle}. In particular, the model predicts a very long solar cycle of about 15 years during the period 1680-1700. Indeed, \cite{Yamaguchia} have observed a very long solar/temperature related cycle during the period 1680-1700 in detailed Japanese tree ring $\delta^{18}O$ data (which are a  proxy for the temperature) and in  $^{14}$C production anomaly (which is a proxy for solar activity).

Figure 6D depicts the modulated three-frequency harmonic model simply superimposed to the global surface temperature (HadCRUT3 available since 1850, http://www.cru.uea.ac.uk/), and the annual average ACRIM total solar irradiance (TSI) satellite composite (http://acrim.com/) since 1981 (note that ACRIM-1 experiment began on February/1980 and we prefer to disregard the ACRIM composite data between 1978 and 1980 because based on a poorer satellite record which may have uncorrected degradation problem). The figure highlights that since  the Dalton minimum the Earth's climate has gradually warmed reaching local maxima around 1880, 1940 and 2000 and local minima around 1910 and 1970. Both the quasi 60-year oscillation and the upward trending (regulated by the 115-year cycle and the millennial cycle) since the  1800 and 1915 solar grand minima is schematically reproduced by the proposed harmonic solar/planetary model. The ACRIM TSI record peaked around 2001 as the model reproduces. By taking into account also a possible anthropogenic warming component, we can expect a steady-to-cooling global climate until around 2030-2035 (Scafetta, in press) as shown in the yellow area in the figure.

Figure 7 shows the modulated three-frequency model, Eq. \ref{Eq555}, against a multi-proxy 5-year resolved reconstruction of the Northern Hemisphere surface temperature \citep{Ljungqvist} during the period 0-2000 A.D..  A common quasi-millennial cycle is evident. These long cycles are made of  well-known historical periods such as: 1-300 A.D., Roman Warm Period (RWP); 400-800 A.D., Dark Age Cold Period (DACP); 900-1300 A.D., Medieval Warm Period (MWP); 1400-1800 A.D., Little Ice Age (LIA); 1900-present, Current Warm Period (CWP). Moreover, there exists   a very good correlation and time matching between the secular oscillations reconstructed by the beats of the harmonic solar/planetary model (period $P_{12}\approx114.783~yr$) and the equivalent secular oscillations observed in the temperature multi-proxy model. For both record it is possible to count 17 cycles covering 1950 years, which are in very good phase agreement. The relatively good correlation is highlighted at the bottom of the figure where the $P_{12}$-year beat harmonic (see Eq. \ref{Eq15}) is compared to a 165-year smooth moving average residual of the temperature signal: the  correlation between the temperature residual and the modeled  115-year harmonic curve is very high ($r_0=0.3$ for 200 points, $P(|r|\geq r_0)<0.1\%$): this result too is  quite remarkable. Slight apparent differences of phases, such as around 700 A.D., may also be due to some error in the proxy temperature model. The 115-year beat cycle, which regulates the secular solar cycles,  was at its minimum in 1923, at its maximum in 1980 and will reach a new minimum in 2038 A.D. within an error of just a few years as discussed above.

  \begin{figure*}[t]
\includegraphics[angle=-90,width=40pc]{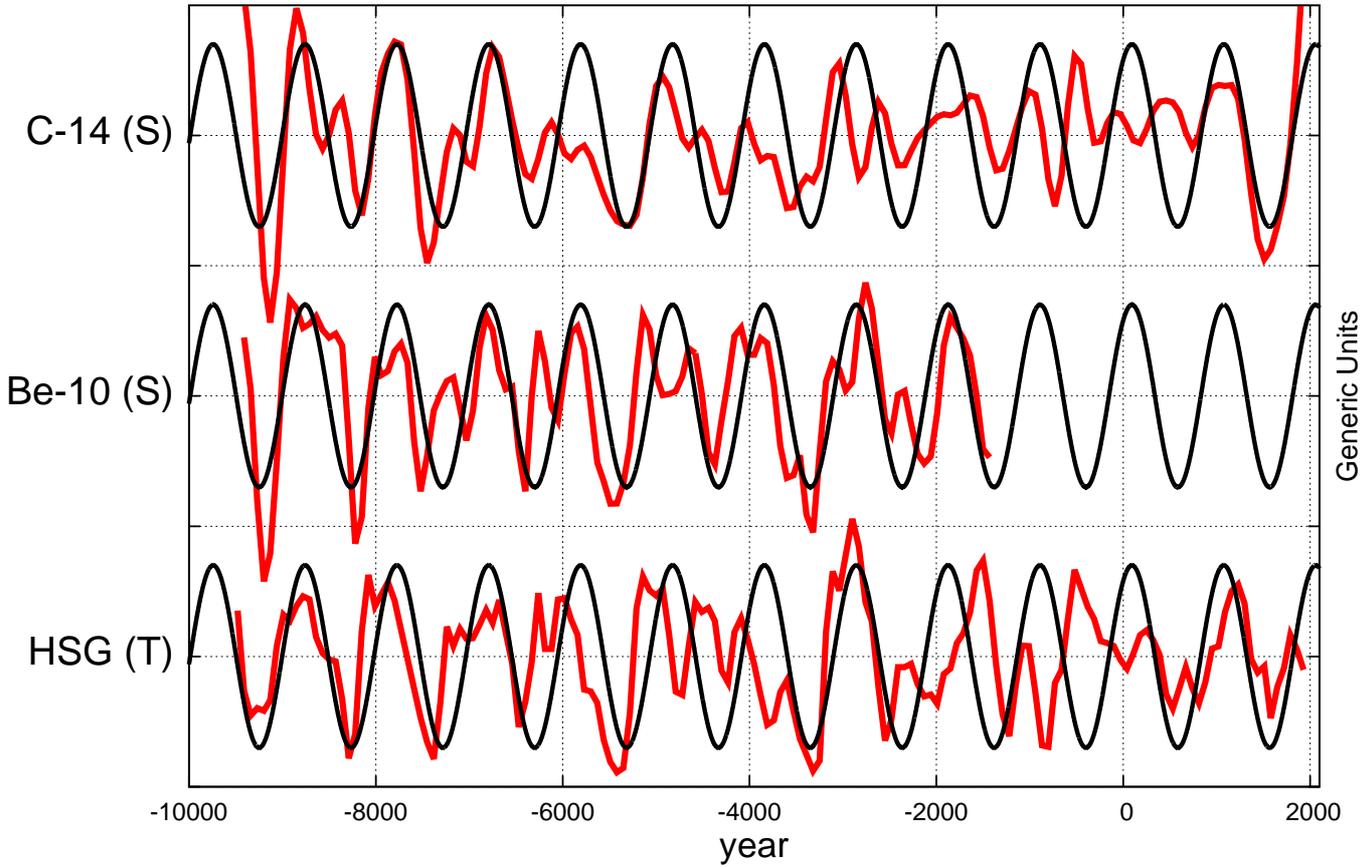}
 \caption{Comparison between carbon-14 ($^{14}$C) and beryllium-10 ($^{10}$Be) nucleotide records, which are used as proxies for the solar activity, and a composite (HSG MC52-VM29-191) of a set of drift ice-index records, which is used as a proxy for the global surface temperature throughout the Holocene \citep{Bond}. The three black harmonic components represent Eq. \ref{eq19} of the solar/planetary millennial beat modulation whose phase, $P_{123}=2059.7$ A.D., is calculated using Eq. \ref{9999}. Note the relatively good correlation between the three solar/climate records and the modeled harmonic function. The correlation coefficients are: $^{14}$C record, $r_0=0.43$ for 164 data ($P(|r|\geq r_0)<0.1\%$);  $HSG$ record, $r_0=0.30$ for 164 data ($P(|r|\geq r_0)<0.1\%$); $^{10}$Be  record, $r_0=0.57$ for 115 data ($P(|r|\geq r_0)<0.1\%$). The symbol ``(S)'' means ``solar proxy'' and ``(T)'' means ``temperature proxy.'' Note that we are using the filtered data records prepared by \cite{Bond} where the very low frequencies at periods $>1800$ years are removed with a Gaussian filter.}
 \end{figure*}

In Figure 8 we compare the prediction of the proposed three-frequency solar/planetary model against the solar and temperature reconstructions during the Holocene, that is, during the last 12,000 years. In these specific cases only the existence and timing of a millennial cycle may be reasonably tested because  long nucleotide and temperature proxy models are characterized by large errors, both in timing and in amplitude \citep{Bond,Bender}, and smaller time-scale patterns may be disrupted. \cite{Bond} compared carbon-14 ($^{14}$C) and beryllium-10 ($^{10}$Be) nucleotides records, which are used as proxies for the solar activity, and a set of drift ice-index records, which were used to construct a proxy for the global temperature throughout the Holocene.
\cite{Bond} concluded that these three records  well correlate to each other and present a common and large quasi-millennial cycle with period of 900 to 1100 years, as we have also found in Figure 4 for other solar proxy models. This millennial frequency band perfectly agrees with the prediction of the three frequency solar/planetary model whose millennial beat cycle ranges between 843-1180 years. We need to test whether the quasi millennial cycles found in the proxy solar/climate records are in phase with the modeled quasi-millennial cycle.

  Figure 8 shows the three records adopted by \cite{Bond} in their figure 3, against the solar/planetary quasi-millennial harmonic cycle with periods of $P_{123}=983.401$ and phase $T_{123}=2059.7$, see Eq. \ref{eq19}.
 Figure 8 clearly shows a good phase matching between the three records and the 983-year harmonic function produced by the solar/planetary model. This demonstrates that the quasi-millennial beat frequency of the three frequency solar/planetary model well reconstructs the millennial cycles of solar and climate proxy records throughout the Holocene. The matching is particularly strong for the $^{10}$Be record for the entire depicted period covering 8,000 years. About the other two records, the millennial cycle matching is quite good almost always, but there appears to be some disruption from 1700 B.C to 200 A.D.. However, in Figure 7 a good matching from 0 A.D. to 2000 A.D., exists for another higher resolved climate reconstruction. So, the poor matching observed from 1700 B.C to 200 A.D. may also be due to some errors in the Bond's data, or to other not well understood phenomena. Indeed, a clearer regular quasi-1000-year cycle since 2000 B.C. is found in ice core proxy temperature records  \citep{Dansgaard,Schulz}. The correlation coefficient between the millennial harmonic function and the other three records are: $^{14}$C record, $r_0=0.43$ for 164 data ($P(|r|\geq r_0)<0.1\%$);  $HSG$ record, $r_0=0.30$ for 164 data ($P(|r|\geq r_0)<0.1\%$); $^{10}$Be  record, $r_0=0.57$ for 115 data ($P(|r|\geq r_0)<0.1\%$). These correlations are highly significant.

  \begin{figure*}[t]
\includegraphics[angle=-90,width=40pc]{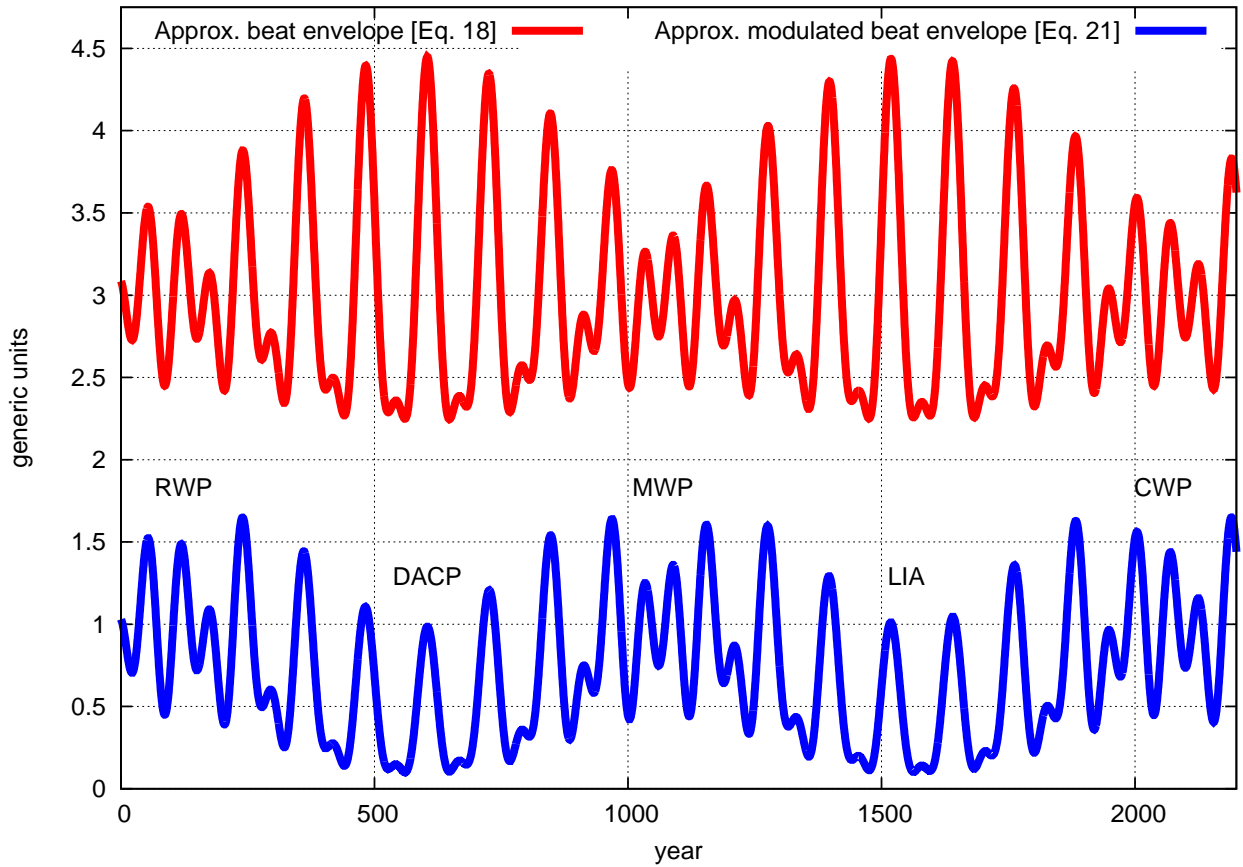}
 \caption{Proposed solar harmonic reconstructions based of four beat frequencies. (Top) Average beat envelope function of the model (Eq. \ref{Eq18}) and (Bottom) the version modulated with a millennial cycle (Eq. \ref{Eq21}). The curves may approximately represent an estimate average harmonic component function of solar activity both in luminosity and magnetic activity. The warm and cold periods of the Earth history are indicated  as in Figure 7. Note that the amplitudes of the constituent harmonics are not optimized and can be adjusted for alternative scenarios. However, the bottom curve approximately reproduces the patterns observed in the proxy solar models depicted in Figure 5. The latter record may be considered a realistic, although schematic, representation of solar dynamics.}
 \end{figure*}

Figure 9 shows an approximate reconstruction of the envelope function of the model (Eq. \ref{Eq18}) and the version modulated with a millennial cycle slight different than that used above (see the Appendix, Eq. \ref{Eq21}). The two depicted curves may approximately represent the average harmonic component of solar activity both in luminosity and magnetic activity, and the bottom one may be quite realistic by approximately reproducing the amplitudes of the solar proxies depicted in Figure 5. Unfortunately, an accurate estimate of the amplitudes of the harmonics is not possible because of the large uncertainty that exists among the solar proxies. The proposed relative amplitudes are roughly calibrated on the solar records depicted in Figure 5, but other solar observable may be characterized by different relative weights. The figure well highlights that solar dynamics is characterized by  quasi-millennial large cycles, by large 115-year cycles and by 60-year smaller cycles that become particularly evident during the millennial maxima.

Note that if Eq. \ref{Eq18} or Eq. \ref{Eq21} is used as proxy solar forcing functions of the climate system, the large heat capacity of the atmosphere/ocean system would respond in a non-linear way  smooth out the higher frequencies and stress the lower ones such as the quasi-millennial cycle. Thus, the depicted curves give only an approximate idea of how the climate on the Earth could have been evolved during the last 2000 years. We also remind that the real solar and climate systems  also depend on other harmonics not taken into account in this work. For example, the  climate records also present super and sub-harmonics of the herein studied cycles, as well as multi-millennial cycles regulated by axial precession, obliquity and eccentricity orbital cycles, which are  known as the Milankovitch's cycles.

\section{Rebuttal of the critique by Smythe and Eddy (1977)}

The empirical results presented above clearly support a theory of planetary influence on solar dynamics. In fact, all major observed solar activity and climate modulations from the 11-year Schwabe solar cycle to the quasi-millennial cycle throughout the Holocene can be reasonably well reproduced by a model based on the frequencies and timings of the two major Jupiter and Saturn planetary tides, plus a median Schwabe solar cycle with a 10.87-year period. Of course, the usage of a larger number of accurate frequencies would probably improve the performance of the model. For comparison, 35-40 solar/lunar planetary harmonics are currently used for reconstructing and forecasting the ocean tides \citep{Thomson,Ehret}. But we leave this exercise and other possible optimizations to other  studies.

We believe that our result is already quite important because it is sufficient to  rebut a fundamental critique  proposed by \cite{Smythe}. These authors argued that Jupiter and Saturn's planetary tides during the solar Maunder minimum present patterns indistinguishable from those observed in the modern era. Thus, they concluded that planetary tides could not influence solar dynamics. Essentially, planetary tidal patterns were found uncorrelated   with the solar records, which present complex patterns such as the Oort, Wolf, Sp\"orer, Maunder and Dalton grand solar minima plus the quasi millennial cycle, etc.

The above critique in the early 1980s definitely convinced most solar scientists to abandon the planetary theory of solar variation first proposed in the 19$^{th}$ century. In fact, while good correlation patterns stimulate researchers to look for a possible explanation and their physical mechanisms, a lack of correlation can be easily interpreted as if no physical link truly exists!

On the contrary, we have found excellent correlation patterns and demonstrated  that the observed decadal, multi-decadal, secular and millennial solar cycles  can easily emerge as interference patterns between the two major tidal cycles induced by Jupiter and Saturn plus an internal solar dynamo cycle, which in our model is indicated by the 10.87 year central Schwabe harmonic. The 10.87-year period is about the average between the 9.93-year and 11.86-year tidal periods, suggesting that a central internal solar dynamo cycle may be approximately    synchronized to and resonating with the average beat tidal period of Jupiter and Saturn.

Smythe and Eddy made the mistake of not taking into account also the fact that solar variations had to be the result of a \emph{coupling} between internal solar dynamo dynamics and external planetary tidal forcing, not just of the planetary tides alone: when the planetary tides and the internal harmonic solar mechanisms they activate interfere destructively with the internal solar dynamo cycle, the Sun becomes quieter and periods such as the solar Maunder minimum occur, as Figures 5-8 show. Thus, we conclude that a solar dynamo theory and a planetary-tidal theory of solar variation are complementary, not in opposition: there is the need of both of them to understand the evolution of solar dynamics!

Of course, herein we are not rebutting other classical objections to the planetary/solar theory. However, these objections are based on classical physical arguments that may simply reveal our current ignorance of the physics involved in the phenomenon. For example, \cite{Jager} further developed the classical argument that planetary tidal elongation on the Sun is tiny. However, this critique simply requires the existence of a strong amplification feedback mechanism that may be provided by a tidal stimulation of the nuclear fusion rate   \citep{Wolff}, which perhaps may be also helped by collective synchronization resonance effects \citep{Pikovsky,Scafetta,Scafetta2011}.  Another objection is based on the Kelvin-Helmholtz timescale and claims that the very slow diffusion of photons in the solar radiative zone would smooth out any luminosity variation signal occurring in the core before the signal could reach the convective zone. The solution of this problem simply requires a fast energy-transportation  mechanism.   Indeed, if the planetary tides are modulating solar fusion rate, the solar interior would not be in thermal equilibrium, but would oscillate, and, therefore, produce waves that can propagate very fast inside the Sun. For example, energy could be transferred into solar g-mode oscillations which would provide rapid upward energy information transport \citep{Grandpierre,WolffM,Wolff}. For example, small and very slow periodic changes in the solar luminosity core production could cause a slight periodic modulation in the amplitude and frequency of the internal solar g-wave oscillations, which may evolve as: $G(t)=(1+a \cos(\omega_p~t)) \cos\{[1+ b \cos(\omega_p~t)]\omega_g~t\}$, where $\omega_g$ is the average frequency of the g-waves, $\omega_p$ is the planetary induced harmonic modulation, and $a$ and $b$ are two small parameters. A wave propagation mechanism would transfer the information of the slow luminosity oscillation that  is occurring in the solar core to the  convective zone quite fast.     As \cite{Wolff} argued: \emph{``an event deep in the Sun that affects the nuclear burning rate will change the amount of energy going into the g-mode oscillations. Some information of this is transported rather promptly by g-modes to the base of the Sun's convective envelope (CE). Once these waves deposit energy there, it is carried to the surface in a few months by extra convection.''} Thus, perhaps, the solar core luminosity anomaly triggered by planetary tides may reach the solar surface in just a few months through still unknown mechanisms. The complex physics of this process may explain why the problem has not been fully solved yet by solar scientists.

  \begin{figure*}[t]
\includegraphics[angle=-90,width=40pc]{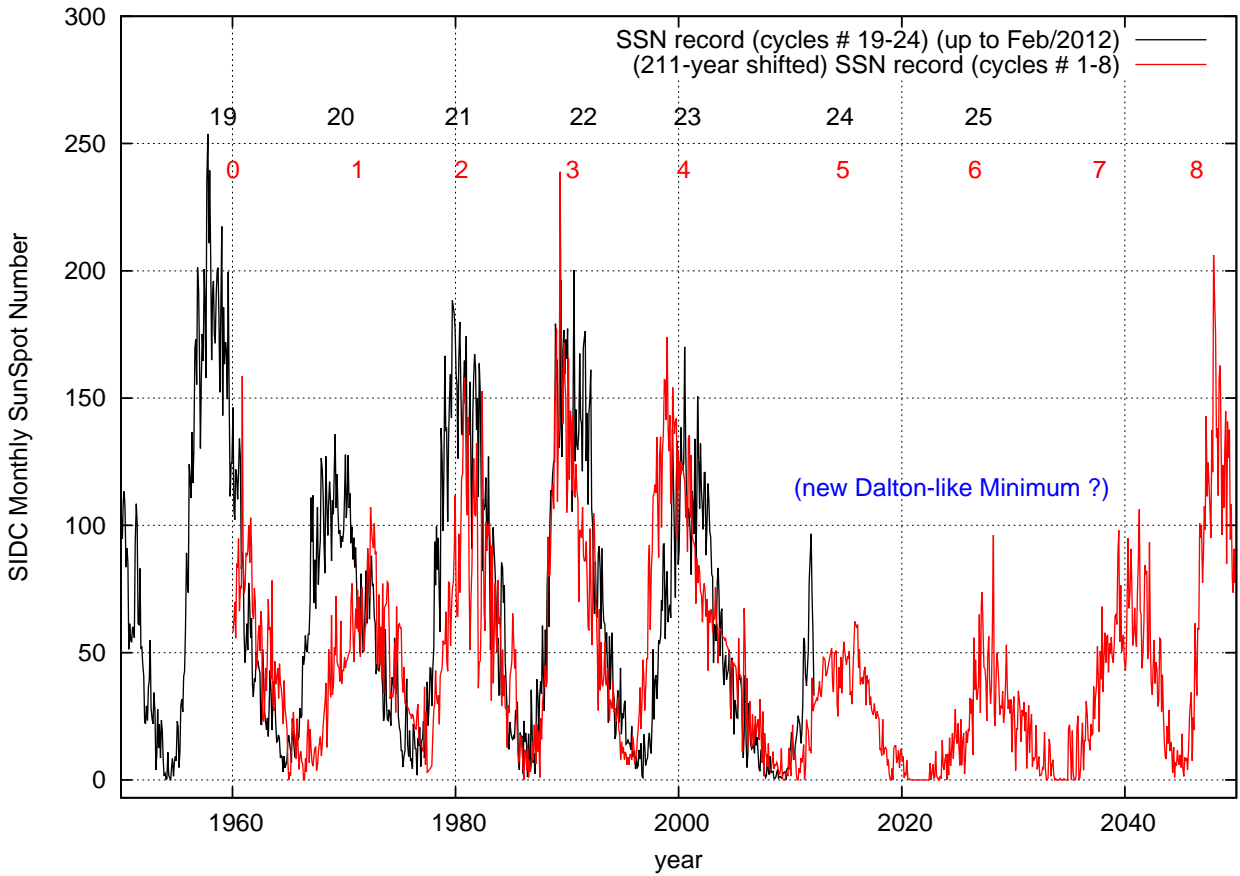}
 \caption{Superposition of the sunspot number cycles 19-23 (1950-2010) and the sunspot number cycles 1-8 (1750-1840). A close similarity between cycles 1-4 and cycles 20-23 is observed. This repeating pattern may suggest that the Sun may manifest a Dalton-like solar minimum during the next few decades, but according Figure 9 the approaching solar minima may be more similar to the minimum around 1910. }
 \end{figure*}

In general, it is perfectly normal in science first to notice good empirical correlations among phenomena, and only  later,  understand the physical mechanisms involved in the processes.
For example, in the 19$^{th}$ century a strong correlation between sunspots and geomagnetic activity  was discovered. However, not only was no physical mechanism  known, but also its own  existence  appeared incompatible with the known physics of the time because the Sun is so far from the Earth that the strength of any solar magnetic field variation is too weak to directly influence the geomagnetic activity on the Earth, as Lord Kelvin calculated \citep{Moldwin}. However, the good correlation between the two phenomena could simply imply a still unknown physical mechanism, and, indeed, the problem was solved later   when it was understood that magnetic field could be transported by injected plasma traveling from the Sun to Earth \citep{Brekke}.
Thus, the fact that it is not known yet  how the planets can influence solar activity, is not an \emph{evidence} that such a thesis is  wrong. Indeed, the demonstrated  good empirical correlation between solar patterns at multiple time-scales for the last 12,000 years and our proposed harmonic model based mostly on the frequencies and phases of Jupiter and Saturn's orbits  would support a planetary influence on the Sun quite convincingly.

\section{Conclusion}

High resolution power spectrum analysis of the sunspot number record since 1749 reveals that the Schwabe frequency band can be split into three major cycles. The periods of the measured cycles are about 10, 11 and 12 years.  The result suggests that the Schwabe solar cycle may be the result of solar dynamo mechanisms constrained by and synchronized to the two major Jupiter and Saturn's tidal frequencies at 9.93-year and 11.86-year. The central cycle is about the average between the two, which also suggests that the Schwabe solar cycle may be produced by collective synchronization of the solar dynamics to the beat average tidal period.

In the paper we have adopted the two planetary tidal frequencies and phases, and deduced a central frequency at $P_2=10.87 \pm 0.01~yr$. We have demonstrated that the resulting three-frequency harmonic model is able to schematically hindcast multiple patterns observed in both solar and temperature records at the decadal, multi-decadal, secular and millennial time-scale throughout the Holocene. The time range used for the hindcast test   is far beyond the 262-year sunspot number period used to estimate some of the free parameters of the harmonic model, while the other free parameters comes from the orbital properties of Jupiter and Saturn. In particular, we have demonstrated that the proposed model reconstructs the timing of the Schwabe solar cycle reasonably well, and hindcasts the timing of the Oort, Wolf, Sp\"orer, Maunder and Dalton solar grand minima and of the quasi-millennial solar cycles very well.
  The model also reconstructs  the timing of quasi  60, 115, 130 and 983-year cycles observed in climate  records, including current global surface temperature records \citep{Scafetta,scafett2011b,scafett2012}. Note that the current paper did not attempt to explicitly estimate the amplitudes of the temperature harmonics induced by the reconstructed solar harmonics: this issue  can be addressed in another  study.

   The model also supports the ACRIM total solar irradiance satellite composite, suggesting that the Sun experienced a secular maximum around 1995-2005, as proposed by \cite{ScafettaW2009}, and reconstructs a local solar maximum around 1940-1450, as shown in some solar proxy models \citep{Hoyt,Christensen,Loehle}. The model also predicts that during prolonged solar minima, the solar cycle length may be statistically longer. For example, the model has hindcast a very long solar cycle of about 15 years during 1680-1700 as observed in both annually-resolved solar and climate proxy records \citep{Yamaguchia}.

  Because the hindcast test has been statistically successful at multiple time scales, the model can be used for forecast purposes too.
  The model predicts that the Sun is rapidly entering into a prolonged period of low activity that may last for a few decades and reach a grand minimum around  2025-2043. This pattern is due to the forecasted 2030-2035 minimum of the quasi 60-year modulation that should be further stressed by the forecasted 2030-2040 minimum of the 115-year cycle.  Indeed, Figure 10 shows that there exists an apparent and approximate cyclical repetition between solar cycles 1, 2, 3 and 4, which occurred just before the Dalton Minimum, and the cycles 20, 21, 22 and 23.

   We observe that the approaching low solar activity may cool the climate slightly more than what calculated in \cite{scafett2012} because there the large 115-year cycle was not taken into account. However, the millennial solar cycle is still slowly increasing and would peak around 2060, while the weaker 130-year cycle would peak in 2035. See Figure 6.

With the best of  our knowledge, none of the current solar theories based on the assumption that the Sun is an isolated system, has been able to hindcast and/or reconstruct  solar dynamics patterns on multiple time scales ranging from the timing of the 11-year Schwabe solar cycle to those of the grand solar minima plus the quasi-millennial solar cycle.

The numerous empirical evidences found in this work clearly rebut the argument made by \cite{Smythe}, and strongly point toward the conclusion that solar-planetary physical coupling mechanisms should exist, as already suggested by multiple authors as discussed in the Introduction. Further research should address the physical mechanisms necessary to integrate planetary tides and solar dynamo physics for a more physically based model.

The Sun is essentially oscillating in the observed way because the cyclical movement of the planets   induces a specific harmonic gravitational forcing on it, and this external forcing couples with the internal nuclear and solar dynamo mechanisms to generate the observed quasi-harmonic solar dynamical evolution. Equivalent harmonics are then forced on the climate system by means of cycles in the solar luminosity and in the solar magnetic forcing of the cosmic ray flux \citep{scafett2011b,scafett2012}. Oscillations of the cosmic ray should contribute to  regulate cloud cover and cause albedo oscillations, which ultimately induce temperature/circulation oscillations in the atmosphere and in the oceans that induce climate changes.  Thus, there appears to be an indirect link through the Sun between the Earth's climate and the geometry of the planetary motion, as it was commonly believed by all ancient scientists \citep{Ptolemy,Masar,Kepler} and  advocated by some modern scientists as well, as discussed in the Introduction.

In conclusion, the proposed solar/planetary harmonic model, whose equations are summarized in the Appendix, has well captured all major harmonic feature observed in numerous solar and climate records throughout the Holocene at multiple temporal scales. Therefore, it may be used for partial solar and climate forecast purpose too.
This study has identified secular and millennial natural harmonics at 115, 130 and 983 years, which need to be added to the 9.1, 10-11, 20 and 60 year solar/planetary/lunar harmonics already identified in other studies \citep{Scafetta,scafett2011b}. Note that the 115- and the 983-year cycles were in their warming phase for most of the 20$^{th}$ century (see Figures 7 and 8) and, together with the 60-year modulation, could have significantly  contributed to the observed warming secular trend and global surface temperature modulation of the 20$^{th}$ century. This result would be consistent with our previous empirical finding claiming that up to 70\% of the observed post-1850 climate change and warming could be associated to multiple solar cycles poorly processed by current general circulation models \cite{Scafetta2007,Scafetta2009,scafett2012}.

\section*{Appendix A. Model equations and supplementary data}

Here we summarize the functions used for constructing the planetary/solar harmonic model in generic relative units.

The three basic proposed harmonics are:

\begin{equation}\label{}
    h_1(t)=0.83~\cos\left(2\pi ~\frac{t-2000.475}{9.929656}\right)
\end{equation}

\begin{equation}\label{}
    h_2(t)=1.0~\cos\left(2\pi ~\frac{t-2002.364}{10.87}\right)
\end{equation}

\begin{equation}\label{}
    h_3(t)=0.55~\cos\left(2\pi ~\frac{t-1999.381}{11.862242}\right)~,
\end{equation}
where the relative amplitudes are weighted on the sunspot number record. The basic harmonic model is

\begin{equation}\label{Eq13}
    h_{123}(t)=h_1(t)+h_2(t)+h_3(t)
\end{equation}

\begin{equation}\label{Eq14}
    f_{123}(t)=
\begin{cases}
  h_{123}(t)  & \text{if } h_{123}(t)\geq 0 \\
  0 & \text{if } h_{123}(t)<0
\end{cases}
\end{equation}
which is depicted in Figures 5A.
The chosen beat function modulations in generic relative units and their sum are:

\begin{equation}\label{Eq15}
    b_{12}(t)=0.60~\cos\left(2\pi ~\frac{t-1980.528}{114.783}\right)
\end{equation}

\begin{equation}\label{}
    b_{13}(t)=0.40~\cos\left(2\pi ~\frac{t-2067.044}{60.9484}\right)
\end{equation}

\begin{equation}\label{}
    b_{23}(t)=0.45~\cos\left(2\pi ~\frac{t-2035.043}{129.951}\right)
\end{equation}

\begin{equation}\label{Eq18}
    b_{123}(t)=b_{12}(t)+b_{13}(t)+b_{23}(t)+1 ~.
\end{equation}
The three relative amplitudes are roughly estimated against Eq. \ref{Eq14}.
The millennial modulating function is

\begin{equation}\label{eq19}
    g_m(t)=A~\cos\left(2\pi ~\frac{t-2059.686}{983.401}\right)+B
\end{equation}
The parameters $A$ and $B$ may be changed according to the application. The two proposed modulated solar/planetary functions are

\begin{equation}\label{eq20}
    F_{123}(t)=g_m(t)~f_{123}(t)~~~\texttt{with}~~A=0.2, B=0.8
\end{equation}

\begin{equation}\label{Eq21}
    B_{123}(t)=g_m(t)~b_{123}(t)~~~\texttt{with}~~A=0.3, B=0.7~.
\end{equation}
Eq. \ref{eq20} is depicted in Figures 5B and 6.  Eq. \ref{Eq18} and Eq. \ref{Eq21} are depicted in Figure 9.

A supplementary data file can be found in the online version of the paper. It contains 6-monthly values of the functions $f_{123}(t)$, $F_{123}(t)$, $b_{123}(t)$ and $B_{123}(t)$ from 10000 B.C. to 3000 A.D..\newline

\section*{Appendix B. Supplementary data}
Supplementary data associated with this article can be found
in the on line version at\newline\newline
\url{http://dx.doi.org/10.1016/j.jastp.2012.02.016}

\section*{Acknowlegment:}
 The author thanks the ACRIMSAT/ACRIM3 Science Team for support.

%

%

%

%
%

%
%
%

\newpage

\end{document}